\documentclass[10pt,twocolumn,letterpaper]{article}

\usepackage{cvpr}              

\usepackage{graphicx}
\usepackage{amsmath}
\usepackage{amssymb}
\usepackage{booktabs}

\usepackage{times}
\usepackage{epsfig}
\usepackage{subcaption}
\usepackage{acro}
\DeclareAcronym{MRI}{
short=MRI,
long=Magnetic Resonance Imaging
}
\DeclareAcronym{CT}{
short=CT,
long= Computed Tomography
}
\DeclareAcronym{CS}{
short=CS,
long=Compressed Sensing
}
\DeclareAcronym{PI}{
short=PI,
long=Parallel Imaging
}
\DeclareAcronym{CNN}{
short=CNN,
long=Convolutional Neural Network
}
\DeclareAcronym{TR}{
short=TR,
long=Repetition Time
}
\DeclareAcronym{TE}{
short=TE,
long=Echo Time
}
\DeclareAcronym{T1WI}{
short=T1WI,
long=T1 Weighted Imaging
}
\DeclareAcronym{T2WI}{
short=T2WI,
long=T2 Weighted Imaging
}
\DeclareAcronym{FLAIR}{
short=FLAIR,
long=Fluid-Attenuated Inversion Recovery
}
\DeclareAcronym{FT}{
short=FT,
long=Fourier Transform
}
\DeclareAcronym{WCE}{
short=WCE,
long=Wireless Capsule Endoscopy
}
\DeclareAcronym{IFT}{
short=IFT,
long=Inverse Fourier Transform
}

\def\@fnsymbol#1{\ensuremath{\ifcase#1\or *\or \dagger\or \ddagger\or
   \mathsection\or \mathparagraph\or \|\or **\or \dagger\dagger
   \or \ddagger\ddagger \else\@ctrerr\fi}}
\newcommand{\ssymbol}[1]{^{\@fnsymbol{#1}}}

\usepackage[pagebackref,breaklinks,colorlinks]{hyperref}

\usepackage[capitalize]{cleveref}
\usepackage{graphicx}
\usepackage{float}
\usepackage{multirow}
\crefname{section}{Sec.}{Secs.}
\Crefname{section}{Section}{Sections}
\Crefname{table}{Table}{Tables}
\crefname{table}{Tab.}{Tabs.}

\def\cvprPaperID{2578} 
\def\confName{CVPR}
\def\confYear{2022}

\begin{document}

\title{Gastrointestinal Polyps and Tumors Detection Based on Multi-scale Feature-fusion with WCE Sequences}

\author{Falin Zhuo\textsuperscript{1}, Haihua Liu\textsuperscript{1}, Ning Pan\textsuperscript{1} \\
\textsuperscript{1}South Central University for Nationalities   \hspace{10 mm}  \textsuperscript{2}Nanjing Eastern Theater General Hospital   \hspace{10 mm} \\
}

\maketitle

\begin{abstract}
Wireless Capsule Endoscopy(WCE) has been widely used for the screening of gastrointestinal(GI) diseases, especially the small intestine, due to its advantages of non-invasive and painless imaging of the entire digestive tract.However, the huge amount of image data captured by WCE makes manual reading a process that requires a huge amount of tasks and can easily lead to missed detection and false detection of lesions.Therefore, In this paper, we propose a \textbf{T}wo-stage \textbf{M}ulti-scale \textbf{F}eature-fusion learning network(\textbf{TMFNet}) to automatically detect small intestinal polyps and tumors in WCE image sequences. Specifically, TMFNet consists of lesion detection network and lesion identification network. Among them, the former improves the feature extraction module and detection module based on the traditional Faster R-CNN network, and readjusts the parameters of the anchor in the region proposal network(RPN) module;the latter combines residual structure and feature pyramid structure are used to build a small intestinal lesion recognition network based on feature fusion, for reducing the false positive rate of the former and improve the overall accuracy.We used 22,335 WCE images in the experiment, with a total of 123,092 lesion regions used to train the detection framework of this paper. In the experiment, the detection framework is trained and tested on the real WCE image dataset provided by the hospital gastroenterology department. The sensitivity, false positive and accuracy of the final model on the RPM are 98.81$\%$, 7.43$\%$ and 92.57$\%$, respectively.Meanwhile,the corresponding results on the lesion images were 98.75$\%$, 5.62$\%$ and 94.39$\%$. The algorithm model proposed in this paper is obviously superior to other detection algorithms in detection effect and performance.


\end{abstract}

\begin{figure*}[htb!]
\centering
\includegraphics[width=1.00\textwidth]{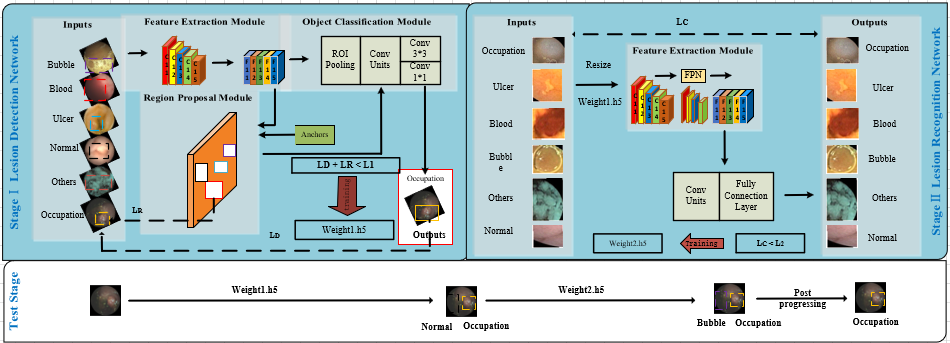}
\caption{An overview of the \textbf{T}wo-stage \textbf{M}util-scale \textbf{F}eature-fusion \textbf{N}etwork (\textbf{TMFNet}).
The detection of WCE lesions by TMFNet is mainly divided into two parts: training part and testing part of the cascade model. The training part includes lesion detection\ref{fig:detection} and lesion recognition\ref{fig:recognition}. In the lesion detection stage, the Faster R-CNN network initialized with ImageNet weights is used to train the WCE image sequence with location information and lesion category information to form the best network weight Weight1.h5 (\textbf{upper} \textbf{left}); the lesion recognition stage is initialized with the detection stage weight Weight1.h5. The recognition network with FPN is used to train the accurately cut lesion images with category labels, and the best network weight Weight2.h5 (\textbf{upper} \textbf{right}) is saved. Finally, the weights of lesion detection and recognition are loaded in turn for the detection of test images, and output The location and category of the lesion. It can be seen from the flow chart that after the detection stage, the preliminary accurate target position can be obtained, but there is false positive target information; however, after the recognition stage, the target category can be corrected and accurate, and the final post-processing retains the location and category information of the lesion. The whole process is end-to-end and requires only one use of the non-maximal suppression(NMS) algorithm at any stage of detection and recognition. Occupation in the figure specifically refers to the tumor and polyp lesions studied in this paper. The loss ($\mathcal{L}_{R}$) and loss ($\mathcal{L}_{D}$) refer to the loss of the RPM and the target detection module in the Faster R-CNN network respectively. The lower limit of the sum of the two losses is ($\mathcal{L}_{1}$) meanwhile ($\mathcal{L}_{C}$) refers to the loss of the lesion recognition network, the lower bound of the loss value is ($\mathcal{L}_{2}$). When testing the WCE sequence(\textbf{last line}), the weights will be loaded into the respective networks, and finally the location and category of the lesions will be drawn on the original image.}
\label{fig:pipline}
\end{figure*}

\section{Introduction}
Wireless Capsule Endoscopy (WCE) has been highly praised by doctors and patients in clinic since its birth, mainly for the following reasons: (1) WCE is simple and painless to operate, requiring neither anesthesia and other drugs, nor It can avoid the risks of gastrointestinal perforation and cross-infection brought by traditional gastroenteroscopy. (2) So far, almost no traditional push-type endoscope can be used alone to complete the detection of the entire gastrointestinal tract of patients. WCE can clearly capture images of the whole digestive tract. (3) WCE is suitable for a wide range of people and has relatively loose requirements. It is suitable for people with complete digestive tract functions and has low preoperative requirements. 

Despite its ability to non-invasively examine the entire gastrointestinal tract, WCE has some limitations in gaining widespread clinical use. Specifically, doctors or clinicians with their experience need to spend about 2 hours to review more than 50,000-80,000 images per patient at a time, which is a time-consuming process. Meanwhile, due to abnormal pictures in full WCE image sequences, in order to prevent omission and misdiagnosis in the image screening process, doctors often need to repeat the screening; in addition, considering the physical constitution, digestive tract function and WCE equipment shooting between different patients The difference in effect has also caused an increase in the difficulty of doctors' work. In summary, it is necessary and significant to develop a method that can automatically detect small intestinal lesions in WCE image sequences, which is conducive to quickly discovering the location of digestive tract diseases and determining the type of disease in patients in advance, also improving work efficiency. At the same time, in order to improve the efficiency and accuracy of diagnosis, the Computer-Aided Diagnosis (CAD) system need high requirements.

Typically, an image judged to be abnormal will contain one or more of polyps, tumors, bleeding, ulcer and crohn, along with irrelevant information such as bubbles, food debris, or hookworms. In the main research object of this paper, polyps are vegetations that grow on the surface of the small intestine. Intestinal tumors have a similar raised shape to polyps, but the tumor volume is slightly larger than that of ordinary polyps. At present, scholars have conducted different studies on WCE redundant images (including bubbles, residues, hookworms, and images with high similarity) and lesion images (polyps, tumors, bleeding, ulcer) to varying degrees.

For example, papers~\cite{iakovidis2010reduction,lee2013reducing,fu2012key,karargyris2010three,karargyris2009video} apply traditional methods such as chromaticity space, texture characteristics, and threshold setting to compare the similarity between images to complete similar image redundancy screening; papers~\cite{wang2009computer,li2009computer,li2009texture} use machine learning algorithms (Support Vector Machine, Random Forest) to learn features composed of color, texture, wavelet transform, etc., and then complete the detection of gastrointestinal bleeding lesions; similarly, the paper~\cite{karkanis2003computer,hwang2010polyp,li2011computer,li2011computer,li2012tumor} introduced similar ideas to complete the tumor in WCE. Classification of polyp lesion images and normal images to obtain lesion images. With the accelerated development of hardware and software equipment and the continuous deepening of deep learning in the field of medical images, the paper based on deep learning network algorithm~\cite{chen2016wireless} completed the redundant screening of WCE images by forming a twin neural network; the paper~\cite{jia2016deep,jia2017study,ghosh2018effective} An algorithm based on Convolutional Neural Network (CNN) is proposed to automatically detect and segment the bleeding area in WCE images; the paper~\cite{he2018hookworm} adopts CNN to detect hookworms; the paper~\cite{zhu2015lesion,billah2018gastrointestinal,yuan2017deep,xu2018automatic} employs deep learning schemes such as CNNs, Sparse AutoEncoders (SAE), etc. to complete the detection of polyp pictures and partially redundant images.

In the field of computer vision and medical image processing, the quality of obtaining feature information will greatly affect the generalization strength of algorithm models. In traditional feature learning, the final feature engineering is often characterized by a low degree of automation, complex feature stacking, and unclear feature orientations, which lead to machine learning algorithm models that are prone to missed detections and false detections in practical applications. The deep learning model can build interpretable and scalable feature vectors in the given data set, and at the same time automatically adjust the weight coefficients, which has strong generalization ability and applicability in practical application experiments. However, in the current paper experiments for polyp and gastrointestinal tumor image screening, the accuracy is poor, and it is difficult to apply and promote in actual clinical lesion screening and CAD.

In view of the above research status and shortcomings, we propose a two-stage multi-scale feature-fusion network (TMFNet). Considering that both polyps and tumors appear as prominent bulges in the pictures, the lesions of the two were uniformly marked as mass in the experiment. In this paper, combining clinical scenarios and actual data, the training samples are classified into six categories: normal, space-occupying, bleeding, ulcers, bubbles and residues. The space-occupying includes polyps and tumors. The network model aims to assist clinicians to complete the screening of polyps and tumors, thereby improving the quality and efficiency of gastroenterologists and saving patients' waiting time.

\begin{enumerate}
\item \textit{Large sequence of WCE images for training and testing.} We used 22,335 WCE images in the experiment, with a total of 123,092 lesion regions used to train the detection framework of this paper; 1,946,270 images that did not participate in the training were used as the test set to test the network detection performance. The entire experiment used WCE image sequences of 91 patients, all of which were desensitized, accounting for 14.91$\%$ of the original total data volume (610).

\item \textit{Network weights trained from real clinical dataset.} We trained on the real data set provided by the digestive department of the hospital and obtained reliable network weight parameters; Next, the patient WCE sequence tested using the model parameters (completely separate and independent from the training set). The test results and performance are excellent, and the generalization ability is strong.

\item \textit{End to end two stage lesion detection framework.} Inspired by recent advances in WCE lesion detection method, we propose a two-stage multi-scale feature-fusion network(TMFNet) for polyp and tumor detection in WCE images, by combining lesion detection network\ref{fig:detection} and lesion recognition network\ref{fig:recognition}. 

\end{enumerate}

\section{Related work}
Considering the angle of the image taken by WCE along with the peristalsis of the digestive tract, the location of the lesion on the original image, and the uncertainty of the shape and size, etc., the application of the model will be affected. Secondly, in the practical application of deep learning detection network in the field of medical images, the sensitivity, false positive and accuracy of the model are indispensable evaluation criteria. Therefore, this part mainly introduces the application of image segmentation algorithm in WCE lesion screening, the application of object detection algorithm in medical images, and the definition of evaluation criteria for medical lesion detection.
\subsection{Performance of Image Segmentation Algorithm in WCE Lesion Screening}
Vázquez et al.\cite{vazquez2017benchmark} used the standard FCN\cite{long2015fully} architecture to complete automatic polyp segmentation, and finally achieved a polyp segmentation effect of 56.07$\%$ Intersection over Union (IoU) on the mixed validation set of CVC-ColonDB and CVC-ClinicDB. Next, Brandao et al.\cite{brandao2018towards} reconstructed the FCN by using different backbones (e.g. AlexNet, VGG-16\cite{simonyan2014very} and ResNet\cite{he2016deep}) and evaluated the modified performance on publicly available datasets, finally showing that the ResNet-101-FCN model is effective in polyp detection The detection effect is improved, and the lateral reflection of the network depth can increase the accuracy of the experimental segmentation. Then, in their research results, Wang et al.\cite{wang2018development} mentioned polyp segmentation in colonoscopy images using algorithms including UNet, FCN, SegNet, and a modified ResNet, and trained on the same publicly available dataset and testing, the split IoU increased to 73.91$\%$. At the same time, the SegNet\cite{badrinarayanan2017segnet} network architecture was used by Wang in polyp detection, but no evaluation metrics for segmentation performance were provided. Kang and Gwak \cite{kang2019ensemble} proposed a Mask R-CNN architecture combining transfer learning and ensemble strategies to segment polyps in colonoscopy images, where the weights transferred to colonoscopy images were derived from the publicly available natural image COCO dataset. Finally, this scheme achieves segmentation results of 66.07$\%$ and 69.46$\%$ IoU on the ETIS-LARIB and CVC-Colon datasets, respectively.
\subsection{Application of object detection algorithm in WCE lesion screening}
Unlike the WCE image classification algorithm\cite{zheng2018localisation} that only provides the category of the test object, the object detection algorithm also needs to further provide the location of the test object. The detection algorithm can be divided into one stage (without RPN, such as SSD\cite{liu2016ssd}, YOLO series\cite{redmon2016you,redmon2017yolo9000,redmon2018yolov3,bochkovskiy2020yolov4}, M2Det\cite{zhao2019m2det}, CenterNet\cite{zhou2019objects}, etc.) and two stages (with RPN, such as Fast R-CNN\cite{girshick2015fast}, Faster R-CNN\cite{ren2015faster}, D2Det\cite{cao2020d2det}, etc.) , Trident-Net\cite{li2019scale}, etc.), both are Anchor-Base; while CornerNet\cite{law2018cornernet}, CornerNet-Lite\cite{law2019cornernet}, CenterNet\cite{duan2019centernet} are Anchor-Free. 

Most existing WCE polyp and tumor lesion detection algorithms are usually based on a supervised learning (image label and bounding box annotation) strategy, using the above object detection architecture to locate polyp regions. Lan et al\cite{lan2019deep}. proposed a cascaded proposal network combined with transfer learning to detect five different anomalies, including residues, hemorrhages, bubbles, tumors, and polyps. The method is that the region proposals generated by the multi-region combination method are re-weighted by the region proposal rejection module, and finally a CNN-based detection module is used to predict abnormal bounding boxes and corresponding categories. The overall test results of the experiment showed the advantage of high accuracy in bleeding screening, but the accuracy of screening tumors and polyps was less than 60$\%$. Mo et al\cite{mo2018efficient}. used Faster R-CNN for polyp detection in conventional colonoscopy images and performed well in the paper experiments. In\cite{urban2018deep}, the regression-based detection algorithm YOLO was trained to locate individual polyps present in each conventional colonoscopy frame. Zheng et al\cite{zheng2018localisation}. further performed cross-dataset validation to evaluate the generalization ability of YOLO to detect polyp regions in colorectal images. Unfortunately, because the model parameters trained by white light (WL) image and narrow band (NB) image features used in the paper are difficult to transfer to cross data sets; secondly, the training data set that the model parameters depend on is too limited, making it difficult to obtain the model generalization. In the paper, the author Zhang\cite{zhang2018polyp} uses a deep learning network named ResYOLO to learn and train using the data set of colonoscopy, and uses a time tracker named Efficient Convolution Operators (ECO) to improve the detection results of the deep learning network. High precision and recall. Aoki et al.\cite{aoki2019automatic} trained an SSD (Single Shotmultibox Detector) model using 5,360 WCE images of ulcer disease, and the sensitivity, specificity and accuracy were high in the test sample containing 10,440 small bowel images (440 images with ulcers).

By comprehensively analyzing the research plans and experimental results of the above scholars, two conclusions can be drawn: (1)The weight of combined transfer learning has certain guiding significance in the process of WCE lesion detection and improves the convergence speed of the model; (2)The use of target detection algorithm The scheme for lesion detection is feasible, but WCE lesion detection based on a supervised strategy requires a large amount of data in order to obtain generalizable weights. At the same time, we are honored to read the papers of Lan\cite{lan2019deep}, Zheng\cite{zheng2018localisation}, Yuan\cite{jia2019wireless}, and are deeply inspired to make this paper.\\
\subsection{Definition of Detection and Evaluation Metrics}
The following table is a commonly used confusion matrix. In the experiment, the Sensitivity, specificity and accuracy are commonly used as the performance evaluation of the detection model. For the calculation of these two indicators, the relationship between the true value and the predicted value of the sample must first be clearly identified, which are mainly divided into four categories: TP (True Positives), FP (False Positives), TN (True Negatives), FN (False Negatives). These four relationships can be clearly shown by the confusion matrix, and the subsequent calculation of precision and recall depends on the confusion matrix shown in tab\ref{tab:confusion}. In addition, the judgment of these four types of samples in target detection needs to calculate the IoU between each prediction frame and the real reference frame. Only when the IoU value is greater than the threshold can the sample be judged as a positive sample.
\begin{table}[!htbp] 
    \centering  
    \caption{The confusion matrix involved in the experiment}
    \label{tab:confusion}
    \begin{tabular}{ccc} \toprule
                     & Positive(Predicted) & Negative(Predicted) \\  \midrule
    Positive(Actual) & TP                  & FN                  \\
    Negative(Actual) & FP                  & TN                   \\  \bottomrule
    \end{tabular}
\end{table}

In order to evaluate and compare the performance of different experimental models from the perspective of data ratio, this study will select the evaluation indicators widely used in medical image classification algorithms: accuracy\ref{accuracy}, sensitivity\ref{sensitivity} and specificity\ref{specificity}.

\begin{equation} \label[]{accuracy}
    accuracy = \frac{TP+FN}{TP+TN+FP+FN}
\end{equation}
\begin{equation} \label[]{sensitivity}
    sensitivity = \frac{TP}{TP+FN}
\end{equation}
\begin{equation} \label[]{specificity}
    specificity = \frac{TN}{FP+TN}
\end{equation}

\section{Methods}
\begin{figure}[!tb]
\centering
\includegraphics[width=0.480\textwidth]{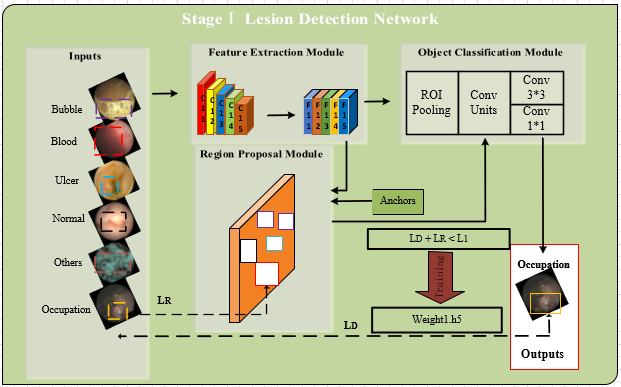}
\caption{The structure of the lesion detection network. Based on the traditional Faster R-CNN, this experiment mainly modified three places: (1) Use the residual structure to replace the VGG structure in the feature extraction module. (2) Modify the parameters of the RPN module, mainly the size of the anchor kernal size. (3) to increase the network depth of the target detection module.}
\label{fig:detection}
\end{figure}

The overall TMFNet pipeline is illustrated in fig\ref{fig:pipline} and consists of two major parts: lesion detection network(fig.\ref{fig:detection}) and lesion recognition network(fig.\ref{fig:recognition}).

\subsection{Lesion detection network}
\noindent\textbf{Modification of Faster R-CNN.} 
The lesion detection network proposed in this study includes a feature extraction module, an RPN model that implements extraction region proposals, and an object detection module that serves as the final labeling and classification of categories. The traditional Faster R-CNN adopts the convolutional neural network VGG16 to implement the feature extraction network in the experiment. Because the VGG structure is prone to gradient disappearance during use, the accuracy of the training phase is terrible, and the model testing phase results in a high rate of missed detection and false detection. Therefore, the residual block ResNet is introduced into the occupancy detection network, and the depth of the network is appropriately increased, so that the model can more accurately label the actual test image without causing the gradient to disappear. Compared with the traditional Faster R-CNN, the following modifications are made in this paper: \textit{(1)}ResNet is used instead of VGG in the feature extraction module (the traditional Faster R-CNN uses the VGG structure). \textit{(2)}The RPN module classification network block is added (the traditional Faster R-CNN is extremely simple to apply the convolution unit RPN). \textit{(3)}ResNet50 is used in the target detection module of Faster R-CNN (traditional Faster R-CNN uses convolutional units with fully connected layers in VGG or ResNet10 to complete the classification function, and the following numbers represent the number of convolutional layers). 

In the Faster R-CNN paper, it is mentioned that the loss function defined by the multi-category detection task is\ref{original faster rcnn loss funcation}:
\begin{equation}
    \begin{aligned}
        \label{original faster rcnn loss funcation}
        L({p_i}, {t_i}) &= \frac{_{1}}{_{N_{cls}}}\sum_{i=1}L_{cls}\left ( p_{i},p_{i^{*}} \right ) \\
        &+ \lambda * \frac{_{1}}{_{N_{reg}}}\sum_{i=1}L_{reg}p_{i}^{*}\left ( t_{i},t_{i^{*}} \right )
    \end{aligned}
\end{equation}
Since the lesion and non-lesion detection involved in this paper can also be classified as multitasking, the meaning of the parameters in the formula is consistent with the original paper. We only redefine the above loss function as the following, which is convenient to correspond to the flow\ref{this paper faster rcnn loss funcation}. In order to solve the class imbalance problem in the experiment, we also increased the focal-loss in the experiment.
\begin{equation}
    \begin{aligned}
        \label{this paper faster rcnn loss funcation}
        L_{D}+L_{R} &= \frac{_{1}}{_{N_{cls}}}\sum_{i=1}L_{cls}\left ( p_{i},p_{i^{*}} \right )+ \frac{_{1}}{_{N_{cls}}}\sum_{i=1}L_{reg}p_{i}^{*}\left ( t_{i},t_{i^{*}} \right ) \\
        &=\frac{_{1}}{_{N_{cls}}}\left ( \sum_{i=1}L_{cls}\left ( p_{i},p_{i^{*}} \right )+\sum_{i=1}L_{reg}p_{i}^{*}\left ( t_{i},t_{i^{*}} \right ) \right )
    \end{aligned}
\end{equation}

\noindent\textbf{Modification of anchors.} 
Object detection is different from image classification. It not only needs to output the category of the object, but also needs to locate the position of the object. The training phase of detection is the process of optimizing and solving the loss function, and the loss function of the target detection model requires the joint participation of the real value and the predicted value. At the same time, the true value and the predicted value are composed of the class label and location label of the target. Therefore, the key to target detection is to make the combination of the corresponding prediction boxes approximate the combination of the true values. The function of the anchor point is to draw candidate boxes of different sizes and ratios on the common feature map. Its parameters are mainly scale and ratio. Using this combination of parameters can The size of the proposed area is determined, and the location information can be calculated by location regression. Although the emergence of Anchor enables the detection algorithm to achieve more accurate detection, it still has shortcomings: 
First, the detector usually judges whether it is a target candidate region based on a set of anchor boxes and a large number of candidate regions, and then return to the effective area, so that the area continuously achieves the maximum overlap with the real area, and using too many Anchors often causes the imbalance of positive and negative samples in the use of the candidate area. At the same time, it slows down the training and testing speed of the model. 
Second, too many anchors will lead to an increase in the hyperparameter learning cost of the network. The hyperparameters of the deep learning network are related to the tasks and data sets of network learning. The learning and tuning of parameters is a difficult task, and the excessive participation of anchors will increase the combination of model parameters. Since the experimental task at this stage is to quickly detect the category and location of the occupant lesions, the use of too complex anchor points will lead to a nonlinear decrease in the detection efficiency of the network. Therefore, the parameter selection of anchor is improved in the experiment. 

The scale of the anchor parameter used by Faster R-CNN is $128^{2}$, $256^{2}$, $512^{2}$ and the aspect ratios are 1:1, 1:2, 2:1. We re-examined the choice of two parameter values in the experiment: (1)Anchor in the RPM module What interval should the parameter scale take to cover the area of the current (large) dataset? Because the region proposal block obtained by the RPN module directly affects the sensitivity and specificity of the target detection module and the entire detection algorithm, the non-maximum suppression algorithm only integrates and divides region proposals of the same category. (2)The anchor parameter in the RPN module Is the ratio really necessary? Because the original Faster R-CNN is slow in practical use, a large part of the detection time is spent on the RPM module. If a quantitatively and qualitatively suitable area proposal is obtained here, the inspection time will be greatly saved. (3)What we need to design is a real-time and accurate lesion detection framework, so how to strike the right balance between time and performance may start in the anchor part. \\
\noindent\textbf{Comparative experiment in this section.} 
In order to effectively highlight the performance advantages of the improved network used in the experiment in the actual test phase, two detection models are introduced in the paper for comparison. 

Imitation Faster R-CNN: Gao and other scholars\cite{gao2019res2net} proposed to use a new residual structure Res2Net with a simple structure and good performance to obtain the scale features in the residual block, and the detection performance on the natural image dataset ImageNet exceeded the residual structure ResNet. Since the scale information of lesions is represented in the WCE image as outline, color, texture and other information, the scale characteristics of lesions are also a potentially effective feature information in practical experiments. Therefore, in the experiment, it is proposed to replace the ResNet block with the Res2Net block to form a new space-occupying detection model, named Imitation Faster R-CNN. 

Faster R-CNN-FPN: Paper\cite{long2015fully} mentioned the FPN structure in the paper. The addition of top-down features can express high-level information more clearly, which is suitable for the extraction and labeling of small targets. Because the deep information contains slightly less semantic features and typical classification features, it is beneficial to the classification of small objects but its location information is lost, but the higher resolution of the shallow information can make up for the location defects of small objects. However, there are small-target polyps in the experimental test data set involved in the paper (Small means that the proportion of the footprint in the original overall image of WCE is less than 15$\%$). The background of deep learning and the needs of actual scenes are integrated, so the detection structure Faster R-CNN-FPN with bidirectional feature fusion is used in the experiment. The feature map obtained after backbone in the traditional Faster R-CNN structure can express the bottom-up feature information of the image, and after adding FPN, it can be recombined into bottom-up and top-down fusion feature information. Multiple sets of 1x1 convolutions used in the fusion stage can transform the number of feature channels and increase the nonlinearity of feature information. Obviously, the feature fusion process also needs to obtain vectors of the same size through upsampling, and complete the splicing of corresponding vectors. Finally, the RPN, pooling operation, detection module and NMS are connected to complete the detection and screening of small intestinal space-occupying lesions based on WCE image sequence.

\subsection{Lesion recognition network}
\noindent\textbf{Background and Purpose.} 
In the stage of lesion detection network alone, the training accuracy is high, the model testing sensitivity is high, the missed detection rate is low, but the false detection rate is too high. There are two reasons for the above phenomenon: First, in the improved Faster R-CNN structure used in the occupancy detection stage, the deep feature information used in classification is the result of deep learning after the feature extraction network. Second, due to the long-short tail effect and data imbalance in the amount of information in each RoI (Region of Interest), the image recognition of the lesion is inaccurate and interferes with the final classification. Therefore, combined with the idea of multi-scale feature fusion, under the premise of ensuring that the sensitivity of lesion detection stage remains unchanged, and with the goal of reducing the false positives of detection, this paper proposes a feature fusion-based WCE sequence digestive tract lesion recognition stage. \\
\begin{figure*}[!tb]
    \centering
    \includegraphics[width=0.480\textwidth]{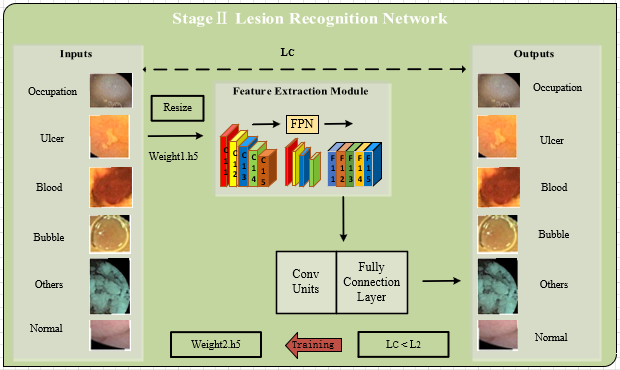}
    \caption{The structure of the lesion recognition network.Due to the high sensitivity of the lesion detection network, but the high false positive rate, this image will affect the efficiency of doctors' screening images and has not solved the fundamental problem. Therefore, this paper proposes to use a lesion identification network to reduce false positives.}
    \label{fig:recognition}
\end{figure*}

\noindent\textbf{Comparative experiment in this section.}
The network structure of occupant lesion recognition proposed in this section (abbreviated as WCE-RFNet) adopts the network structure of ResNet50 as backbone and FPN as neck. The acquisition of multi-scale features in the classification and recognition network based on multi-scale feature-fusion and occupancy candidate region is obtained by the convolutional skeleton backbone, and the fusion method and strategy are completed by FPN. The regional information of the location is to obtain accurate label information to reduce the final false detection rate. In this experiment, the recognition results of WCE-RFNet are compared with the network models of VGG, dual path network DPN (Dual Path Network) and residual network ResNet50. Among them, in the ResNet50 network, the residual block used at this time is the classic residual block structure ResNet; while the DPN network adds atrous convolution, which can increase the range of feature extraction while maintaining a certain number of parameters. Although this study focuses on the detection of small bowel space-occupying lesions, the WCE images of ulcers, bleeding, normal, bubbles and impurities will interfere and misjudgment the binary classification problem of the experimental nature. Therefore, at this time, images other than space-occupying(polyp and tumor) lesions are collectively classified as non-space-occupying(bleeding, ulcers, etc.), and the core binary classification problem remains unchanged.


\section{Experiments}
\subsection{Experimental Settings}
\begin{figure}[H]
    \centering
    \includegraphics[width=\linewidth]{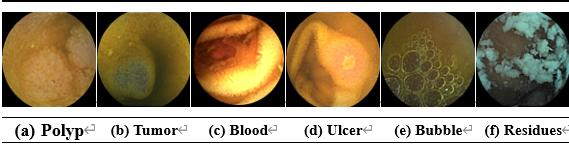}
    \caption{Some abnormal images of digestive tract taken by OMOM WCE, including polyps, tumors, bleeding, ulcers, Images with residues and bubbles.}
    \label{fig:omom_wce_example_images}
\end{figure}

\noindent\textbf{Data description.} 
The data set used in this paper was provided by the Department of Gastroenterology of Nanjing Eastern Theater General Hospital. The OMOM wireless gastrointestinal endoscope developed by Chongqing Jinshan Technology (Group) Co., Ltd. was used. The hospital department collected WCE data from 2009 to 2018. As we all know, in the image sequence captured by WCE, there are often a large number of normal images and redundant images fig\ref{fig:omom_wce_example_images}(e,f), while the proportion of effective lesion images is very small fig\ref{fig:omom_wce_example_images}(a,b,c,d). Therefore, according to the research requirements of this paper and the content of WCE clinical images, in the experiment, the original data set was first labeled as normal, mass, bleeding, and ulcer. Therefore, the original data included 320 normal people, 67 occupants, 56 bleeding and 146 ulcers. In the original data, 14 patients had both bleeding and mass, 7 patients had a diagnosis of bleeding and ulcer, and 1 patient had both bleeding, ulcer and mass. The distribution of specific data can be referred to as shown in tab\ref{tab:omom_wce_ori_database}, which is counted and organized according to whether the number of patients with multiple cases is added, but the data volume of normal labels remains unchanged.
\begin{table}[]
    \caption{Distribution of datasets provided by hospitals.}
    \label{tab:omom_wce_ori_database}
    \begin{tabular}{ccccc} \toprule
                                                               & \begin{tabular}[c]{@{}c@{}}normal \\ (/case)\end{tabular} & \begin{tabular}[c]{@{}c@{}}occupants\\ (/case)\end{tabular} & \begin{tabular}[c]{@{}c@{}}bleeding\\ (/case)\end{tabular} & \begin{tabular}[c]{@{}c@{}}ulcer\\ (/case)\end{tabular} \\ \midrule
    \begin{tabular}[c]{@{}c@{}}without\\ multiple\end{tabular} & 320                                                       & 67                                                          & 56                                                      & 146                                                     \\
    \begin{tabular}[c]{@{}c@{}}with \\ multiple\end{tabular}   & 320                                                       & 82                                                          & 78                                                      & 154                                                    \\ \bottomrule
    \end{tabular}
\end{table}
\noindent\textbf{experimental environment.}
The experimental environment of this paper is an Inspur medical image data analysis server with Centos7 system, Intel(R) Xeon(R) Gold 5118 CPU @ 2.30GHz and NVIDIA Tesla V100-SXM2 graphics card. Among them, the lesion detection network uses the deep learning framework Keras, and the lesion detection network uses the deep learning framework Pytorch. \\
\noindent\textbf{Training and testing of the lesion detection network.}
First, in this stage of the experiment, 20 space-occupying cases, 14 bleeding cases, and 12 ulcer cases were manually selected from the dataset without multiple cases in tab\ref{tab:omom_wce_ori_database} as the source of the training set, and all the remaining cases were as the test set; then, select the corresponding case pictures from the 46 data sequences as the training set, and mark the location and label of the lesion. Because the normal, bubble and residues images in the patient case sequence will interfere with the training of the deep learning network on the occupant images, and the image sequence captured by WCE in the actual clinical scene must also contain these three types of data information, so in the experimental training data During the production of the set, three types of images, normal, bubble-containing and residues-containing, were also selected from 20 space-occupying cases. Next, referring to the standard format of the public natural image data set VOC2007 commonly used in object detection networks, use LabelImg tool to create Own experimental training dataset, which will be called SCUEC-WCE, then store and update the data sequentially according to the time of production. Finally, ask professional doctors and teams to test and correct the accuracy of the experimental dataset.

\begin{table}[]
    \footnotesize
    \centering
    \caption{Arrangement of datasets for WCE training (after dataset augmentation) in lesion detection network experiments}
    \label[]{tab:detection_stage_training_1}
    \resizebox{\linewidth}{!}{
    \begin{tabular}{ccccccc} \toprule
        & space-occupying & bleeding & ulcer & bubble & residues & normal \\ \midrule
        number of patients & 20        & 14    & 12    & 20\textcolor{red}{+}    & 20\textcolor{red}{+}      & 20\textcolor{red}{+}    \\
        number of images   & 5,453\textcolor{red}{*}    & 1,014 & 1,014 & 4,493  & 5,811    & 4,550  \\
        number of object   & 9,226\textcolor{red}{*}    & 9,232 & 1,782 & 17,011 & 35,716   & 50,125 \\
        object/images      & 1.69      & 9.10  & 1.76  & 3.79   & 6.15     & 11.0   \\ \bottomrule
    \end{tabular}
}
\end{table}
Since the process of training phase and testing phase is completely separated when using deep convolutional network. The purpose of the training phase is to let the receptive field of the deep CNNs learn the effective information of the input image, such as edge, color, texture, etc., and convert it into the weight information of each neuron. The test phase applys the loaded weight information. Deep Convolutional Networks, batch tested on real data. Therefore, the data of the experiments in this chapter are also divided into two completely independent training and testing parts. The training set is the training data set SCUEC-WCE, whose arrangement is shown in tab\ref{tab:detection_stage_training_1} below. The last row in the table represents the ratio of the number of manually annotated lesion areas to the number of pictures, \textcolor{red}{*} indicates that the data object is the result after data enhancement, \textcolor{red}{+} indicates that the source of the data is the same, and both are sequences of patients with space-occupying patients.

In addition, during the training of deep convolutional network learning, the training data needs to be divided into training set and validation set again according to a certain proportion. The detailed distribution is shown in tab\ref{tab:detection_stage_training_testing}. The ratio used in this paper in the current experiment is 9:1, that is, 90$\%$ of the number of pictures corresponding to the data in tab\ref{tab:detection_stage_training_1} is used as the training set, and the remaining 10$\%$ is used as the validation set. In the testing phase, 21 cases were used, including 7 occupants, 7 bleeding and ulcer, and 7 normal. Train in the table refers to the data volume of the SCUEC-WCE dataset as a training sample during the training process, Val. refers to the data volume of the verification sample; Train-Val. refers to the training sample and verification sample data volume. Img. represents the number of corresponding pictures contained in the corresponding batch of samples, and Obj. refers to the number of categories contained in the sample. It can be seen from Table 4 that the paper used 22,335 WCE images in the training of this stage of the experiment, with a total of 123,092 lesion areas. The Test field at the end of the table refers to the data volume of images and mass lesions of the test sample. The WCE image sequence of these 7 patients contains 2,923 space-occupying images, which can be annotated with 3,070 lesion areas.

\noindent\textbf{Training and testing of the lesion recognition network.}
In this stage, the training data used at this time is based on the training data set SCUEC-WCE to extract the ROIs of the image, including occupants, bleeding, ulcers, residues, bubbles and normal. The region of interest and corresponding labels are made into the training dataset SCUEC-WCE-II of this chapter. Next, the training data set SCUEC-WCE-II is sent to the classification and recognition network initialized by the detection weights for learning, and the model weight file is saved after the training. The dataset in the testing phase is the same as the dataset used in the testing phase of the lesion recognition phase. In the testing phase, the region of interest obtained by the RPM is input into the network loaded with recognition weights to output accurate labels, and finally use the NMS algorithm to fuse the positions of the same labels, as shown as fig\ref{fig:recognition}. 

Although the data set at this stage is still an extension of the data in tab\ref{tab:detection_stage_training_1}, the training images at this time are segmented from the original WCE (240*256 and 256*256, the image pixels of the old and new versions are different) images The area blocks with lesions (tumors, polyps, bleeding, ulcers) and non-lesions (bubbles, impurities, normal) constitute a new training dataset SCUEC-WCE-II. In this experiment, the training data was distributed according to the ratio of 8:2, and the training set and validation set required for the training phase of the occupant lesion identification network were obtained. The detailed distribution is shown in tab\ref{tab:recognition_stage_training_testing}. It can be seen from the table that 123,092 WCE images are used in the training of the lesion recognition stage, which is a supervised learning. Since the test set used in this experiment in the testing phase is the same as the test set in the lesion detection stage, and the number of space-occupying lesions in the training stage is equal to the number of space-occupying lesions, 2,923 space-occupying images can be manually selected from the test data. 3,070 lesion areas were labeled. Therefore, 12,149 images of space-occupying lesions and 12,296 space-occupying lesion areas were finally used in this experiment.

The testing process of the lesion recognition stage initially used the WCE sequences of 21 patients (7 space-occupying, 7 non-space-occupying, and 7 normal case data) in the lesion recognition stage, and then based on the test data set. On the above, data of 7 space-occupying, 9 non-space-occupying and 8 normal cases have been added. Sort out the data content as shown in tab\ref{tab:all_stage_training_testing_add}. The table is based on the changes of the test set before and after the test is added; the statistics mainly include the overall number of cases, the number of occupants cases, the proportion of occupants number, the number of overall pictures, the number of occupants patient pictures, the proportion of occupant pictures, and the proportion of occupant pictures. Statistical data from 8 perspectives, including the number of place annotations and the proportion of occupants labels, among which the proportion of occupant numbers refers to the ratio of the number of occupants to the total number of cases, and the proportion of occupants pictures is calculated as the proportion of occupants lesions. The ratio of the number of pictures to the total number of pictures and the proportion of occupathon annotations represent the ratio of the number of manually annotated space-occupying lesion frames to the number of pictures of patients with space-occupying patients.
\begin{table}[!hbp]
    \footnotesize
    \centering
    \caption{Specific data distribution of training and testing process in the lesion detection network experiment}
    \label[]{tab:detection_stage_training_testing}
    \resizebox{\linewidth}{!}{
    \begin{tabular}{ccccccccccc} \toprule
        \multirow{2}{*}{} & \multicolumn{2}{c}{Train} & \multicolumn{2}{c}{Val.} & \multicolumn{2}{c}{Train-Val.} & \multicolumn{2}{c}{Test} & \multicolumn{2}{c}{Total} \\ \midrule
        & Img.        & Obj.        & Img        & Obj.        & Img           & Obj.           & Img         & Obj.       & Img         & Obj.        \\
        space-occupying         & 4,908       & 8,295       & 545        & 931         & 5,453         & 9,226          & 2923        & 3070       & 8,376       & 12,296      \\
        bleeding             & 913         & 8,308       & 101        & 924         & 1,014         & 9,232          & /           & /          & /           & /           \\
        ulcer             & 913         & 1,607       & 101        & 175         & 1,014         & 1,782          & /           & /          & /           & /           \\
        bubble            & 4,044       & 15,327      & 449        & 1,684       & 4,493         & 17,011         & /           & /          & /           & /           \\
        residues          & 5,230       & 32,165      & 581        & 3,551       & 5,811         & 35,716         & /           & /          & /           & /           \\
        normal            & 4,094       & 45,034      & 456        & 5,091       & 4,550         & 50,125         & /           & /          & /           & /           \\
        total             & 20,102      & 110,736     & 2,233      & 12,356      & 22,335        & 123,092        & /           & /          & /           & /           \\ \bottomrule
    \end{tabular}
}
\end{table}
\begin{table}[]
    \footnotesize
    \centering
    \caption{Specific data distribution of training process in the lesion recognition network experiment}
    \label[]{tab:recognition_stage_training_testing}
    \begin{tabular}{cccc} \toprule
              & Train  & Val.   & Train-Val. \\ \midrule
    space-occupying & 7,381  & 1,845  & 9,226      \\
    bleeding     & 7,466  & 1,766  & 9,232      \\
    ulcer     & 1,426  & 356    & 1,782      \\
    bubble    & 13,609 & 3,402  & 17,011     \\
    residues  & 28,573 & 7,143  & 35,716     \\
    normal    & 40,100 & 10,025 & 50,125     \\
    All       & 98,555 & 25,437 & 123,092    \\ \bottomrule
    \end{tabular}
\end{table}
\begin{table}[!hbp]
    \footnotesize
    \centering
    \caption{The specific data distribution of the training process of the lesion recognition network in the experiment before and after the data increase.}
    \label[]{tab:all_stage_training_testing_add}
    \resizebox{\linewidth}{!}{
    \begin{tabular}{ccccccc} \toprule
        \multirow{2}{*}{}                                                              &             & \multicolumn{2}{c}{Before increase} &           & \multicolumn{2}{c}{After increase} \\ \midrule
        & Train Stage & Test Stage        & All Stage       & Add       & Test Stage       & All Stage       \\
        \begin{tabular}[c]{@{}c@{}}number of\\ overall cases\end{tabular}              & 36          & 21                & 57              & 24        & 45               & 81              \\
        \begin{tabular}[c]{@{}c@{}}number of\\ occupant cases\end{tabular}             & 20          & 7                 & 27              & 7         & 14               & 34              \\
        \begin{tabular}[c]{@{}c@{}}occupant cases\\ /overall cases(\%)\end{tabular}    & 55.56       & 33.33             & 47.36           & 29.17     & 31.11            & 41.98           \\
        \begin{tabular}[c]{@{}c@{}}number of \\ all images\end{tabular}                & 22,335*     & 856,362           & 878,697         & 1,089,908 & 1,946,270        & 1,968,605       \\
        \begin{tabular}[c]{@{}c@{}}number of\\ occupant images\end{tabular}            & 5,453*      & 310,677           & 316,130         & 319,600   & 630,277          & 635,730         \\
        \begin{tabular}[c]{@{}c@{}}occupant images\\ /all images(\%)\end{tabular}      & 24.41       & 36.28             & 35.98           & 29.32     & 32.39            & 32.29           \\
        \begin{tabular}[c]{@{}c@{}}number of \\ occupant object\end{tabular}           & 9,226*      & 3,070             & 12,296          & 2,360     & 5,430            & 14,656          \\
        \begin{tabular}[c]{@{}c@{}}occupant object\\ /occupant images(\%)\end{tabular} & 1.69        & 0.99              & 0.39            & 0.74      & 0.86             & 2.31            \\ \bottomrule
    \end{tabular}
}
\end{table}

\section{Results}
\subsection{lesion detection network}
\noindent\textbf{Modification of Faster R-CNN.}
The training is performed on the same batch of data set SCUEC-WCE before and after the network modification, and the experimental training data is compared in tab\ref{tab:detection_networks_selection} in the stage of lesion detection. The indicator Mean Overlap Boxes is the number of positive sample areas provided by the RPN module; the indicator Classifier Accuracy of RPN is the classification accuracy of the lesion foreground in the RPN; the indicator Loss RPN Classifier is the loss of RPN model classification; the indicator Loss RPN Regression refers to the RPN regression Similarly, Loss Detector Classifier is the classification loss of lesions in the detection module; Loss Detector Regression is the loss of location regression of lesions after passing through the detection module; and Elapsed Time represents the time required to run the training program phase. ResNet40-1-ResNet10 represents a detection network structure composed of 40 layers of ResNet convolution for feature extraction, 1 simple convolution layer to form RPN, and 10 layers of ResNet convolution and a fully connected layer to join; while Res40-1-ResNet50 represents a A 40-layer ResNet block is used for lesion feature extraction, a simple convolutional layer is used to form an RPN, and a detection network structure composed of a 50-layer ResNet convolutional block and a fully connected layer is used at last. Similarly, ResNet40-20-ResNet50 represents a 40-layer ResNet The convolution unit is used for feature extraction, and 20 refers to the detection network structure composed of 50 layers of ResNet convolution layers and fully connected layers with an appropriate number of convolution layers added to the RPN module. The network ResNet40-20-ResNet50\textcolor{red}{*} in the table represents the result of using the detection structure of ResNet40-20-ResNet50 and the number of training times is 500; the rest of the detection network statistics are the corresponding data statistics results when the number of training times is 200. \\
\begin{table}[]
    \footnotesize
    \centering
    \caption{Comparison table of training on the same batch of data before and after the improvement of the detection network.}
    \label[]{tab:detection_networks_selection}
    \resizebox{\linewidth}{!}{
    \begin{tabular}{cccccccc} \toprule
        & \begin{tabular}[c]{@{}c@{}}Mean\\ Overlap\\ Boxes\end{tabular} & \begin{tabular}[c]{@{}c@{}}Classifier\\ Accuracy\\ of RPN\end{tabular} & \begin{tabular}[c]{@{}c@{}}Loss of\\ RPN\\ Classifier\end{tabular} & \begin{tabular}[c]{@{}c@{}}Loss of \\ RPN\\ Regression\end{tabular} & \begin{tabular}[c]{@{}c@{}}Loss of \\ Detector\\ Classifier\end{tabular} & \begin{tabular}[c]{@{}c@{}}Loss of \\ Detector\\ Regression\end{tabular} & \begin{tabular}[c]{@{}c@{}}Elapsed \\ Time\\ (ms)\end{tabular} \\ \midrule
        Vgg13-Vgg3                                                       & 20.3                                                           & 0.77                                                                   & 3.33                                                               & 0.18                                                                & 0.45                                                                     & 0.21                                                                     & 515.3                                                         \\
        \begin{tabular}[c]{@{}c@{}}ResNet40-1\\ -ResNet10\end{tabular}   & 23.5                                                           & 0.83                                                                   & 2.89                                                               & 0.13                                                                & 0.41                                                                     & 0.21                                                                     & 627.5                                                         \\
        \begin{tabular}[c]{@{}c@{}}ResNet40-1\\ -ResNet50\end{tabular}   & 27.9                                                           & 0.93                                                                   & 1.37                                                               & 0.06                                                                & 0.16                                                                     & 0.08                                                                     & 727.3                                                         \\
        \begin{tabular}[c]{@{}c@{}}ResNet40-20\\ -ResNet50\end{tabular}  & 29.6                                                           & 0.94                                                                   & 3.26                                                               & 0.09                                                                & 0.15                                                                     & 0.06                                                                     & 694.1                                                         \\
        \begin{tabular}[c]{@{}c@{}}ResNet40-20\\ -ResNet50\textcolor{red}*\end{tabular} & 27.0                                                           & 0.94                                                                   & 3.15                                                               & 0.07                                                                & 0.15                                                                     & 0.07                                                                     & 705.2                                                         \\ \bottomrule
    \end{tabular}
}
\end{table}
From the analysis in the comparison table, it can be seen that the results of the network before and after the modification in the same batch of experimental training data have been significantly improved. It is mainly reflected in \textit{(1)}the network structure of the RPM using ResNet40-20-ResNet50 with 200 epochs) can provide the largest number of positive sample regions, and the classification accuracy of the border is the highest. \textit{(2)}Due to the modified network depth and The increase of the parameters will lead to the increase of the network learning time. The results in the table are in line with the experimental theoretical expectations, but the index value of the improved structure in the running time is within the acceptable range of the experiment. Therefore, the structure of ResNet40-20-ResNet50 is used in the lesion detection stage.

\begin{figure}[H]
    \centering
    \includegraphics[width=\linewidth]{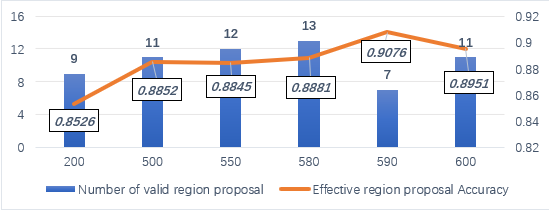}
    \caption{Variation curve of the number of effective region proposal and the accuracy of the improved network in different epochs}
    \label{fig:wce_detection_epoch_selection}
\end{figure}

\begin{figure}[H]
    \centering
    \includegraphics[width=\linewidth]{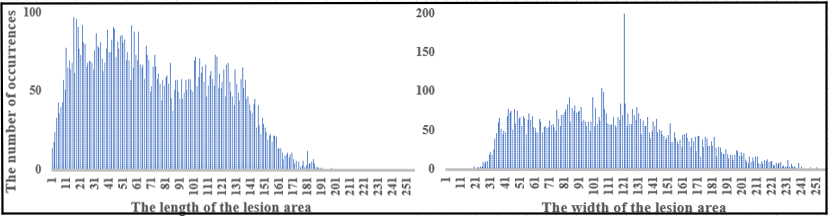}
    \caption{The length and width of the lesion area in the training set}
    \label{fig:wce_length_width_select}
\end{figure}

\begin{figure}[H]
    \centering
    \includegraphics[width=\linewidth]{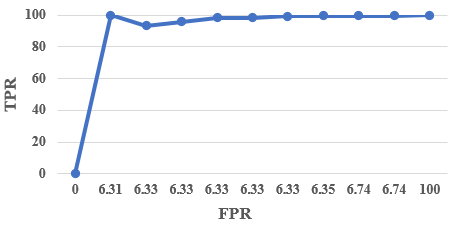}
    \caption{ROC curve for IoU threshold}
    \label{fig:wce_detect_roc}
\end{figure}

\begin{figure}[H]
    \centering
    \includegraphics[width=\linewidth]{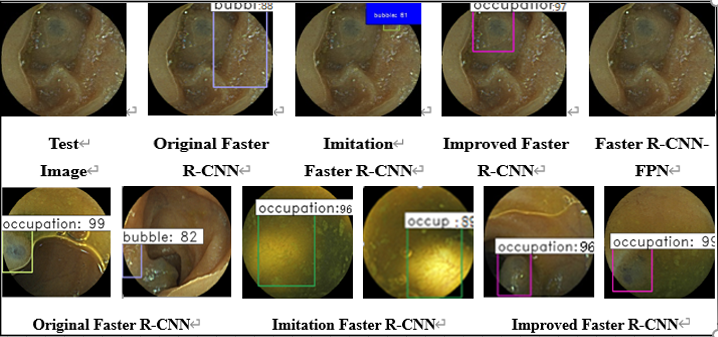}
    \caption{Comparison of the detection and annotation effects of various occupancy detection networks on the same image and different images.}
    \label{fig:wce_detect_result_image}
\end{figure}

After determining the structure of the lesion detection network, this paper draws the change curve of the number of RPNs and the accuracy in different running times in the improved network as shown in fig\ref{fig:wce_detection_epoch_selection}. The number of RPNs refers to the number of overlaps between the suggested area of the occupant lesions obtained after the RPN module in the improved Faster R-CNN network and the lesion area in the real sample, and the RPN accuracy refers to whether the RPN determines whether the current area is the classification accuracy of the foreground or background; the standard of area overlap is (1) The area of the predicted area of the lesion and the labeled area overlap the most. (2) The overlapping area of the predicted area of the lesion and the labeled area accounts for more than 60$\%$ of the total area. \\
It can be seen from the tab\ref{tab:detection_networks_selection} that the detection structure of ResNet40-20-ResNet50 has the largest number of overlaps between the predicted area of the lesion and the labeled area when the number of runs is 580, but the accuracy of classifying foreground and background is slightly lower than that of 590 runs. time accuracy. In the detection phase of this paper, firstly, it is necessary to ensure that the RPN can provide enough suggested areas for lesions, and secondly, the performance of the network structure is also one of the evaluation indicators of the actual model. To sum up, the lesion detection network in this paper finally uses ResNet40-20-ResNet50 to detect the structure, and at the same time determines that the number of hyperparameter runs for network training is 580.
\noindent\textbf{Modification of anchors in RPN.}
The introduction of different scales and aspect ratios through the anchor mechanism in the lesion detection network enables the network to adaptively mark the location of the lesion, but too many or too simple anchors will lead to a decrease in the learning performance of the network, and even reduce the detection effect of the network. Therefore, the improved network is used to conduct experiments and analysis on different combinations of anchor scales and proportions in the same batch of training data set SCUEC-WCE. The statistical results are shown in tab\ref{tab:detection_networks_anchors} below, where 128(1) means that the scale parameter of the anchor at this time is 128, the aspect ratio antio parameter is 1, and so on. The statistical results include two levels of RPN and Image, which represent the acquisition performance of the suggested area of the lesion and the detection performance of the image of the occupied lesion respectively. The statistical data is the overall detection result of the 21 test patients. It can be seen from the table that when the anchor scale and aspect ratio parameters are set to a combination of 64(1) and 128(1), the sensitivity and other indicators are the best, and when the combination of parameters is too complex, it will cause false positives high and low accuracy; secondly, the detection performance brought by complex parameter combinations is lower than the detection performance of simple anchor parameter combinations, and the difference in performance is reflected in the consumption of detection time; however, too simple anchor parameter combinations will lead to overall detection metrics and performance degradation.\\
\begin{table}[!hbp]
    \footnotesize
    \centering
    \caption{Statistical performance of combinations of different Anchor parameters in the testing phase.}
    \label[]{tab:detection_networks_anchors}
    \resizebox{\linewidth}{!}{
    \begin{tabular}{cccccccc} \toprule
        \multirow{2}{*}{\begin{tabular}[c]{@{}c@{}}scale(ratio)\end{tabular}}           & \multicolumn{2}{c}{\begin{tabular}[c]{@{}c@{}}sensitivity\\ (\%)\end{tabular}} & \multicolumn{2}{c}{\begin{tabular}[c]{@{}c@{}}specificity\\ (\%)\end{tabular}} & \multicolumn{2}{c}{\begin{tabular}[c]{@{}c@{}}accuracy\\ (\%)\end{tabular}} & \begin{tabular}[c]{@{}c@{}}elapsed \\ time(ms)\end{tabular} \\ \midrule
        & RPN                                    & Img                                   & RPN                                    & Img                                   & RPN                                  & Img                                  & /                                                           \\
        512(1)                                                                             & 80.6                                   & 78.27                                 & 19.9                                   & 19.3                                  & 80.1                                 & 80.66                                & 210                                                         \\
        128(1)                                                                             & 97.2                                   & 97                                    & 17.3                                   & 16.5                                  & 82.08                                & 82.83                                & 220                                                         \\
        64(1)128(1)                                                                        & \textbf{98.37}                                  & 98                                    & \textbf{6.59}                                   & \textbf{6.3}                                   & \textbf{93.42}                                & \textbf{93.71}                                & 350                                                         \\
        \begin{tabular}[c]{@{}c@{}}64(1,0.5,2)\\ 128(1)\end{tabular}                       & 96.55                                  & 96.48                                 & 7.06                                   & 7.18                                  & 92.55                                & 92.84                                & 400                                                         \\
        \begin{tabular}[c]{@{}c@{}}128(1,0.5,2)\\ 256(1,0.5,2)\\ 512(1,0.5,2)\end{tabular} & 99.8                                   & 100                                   & 13.46                                  & 12.7                                  & 86.64                                & 88.40                                & 751                                                         \\ \bottomrule
    \end{tabular}
}
\end{table}
Fig\ref{fig:wce_length_width_select} plots the length and width distributions of the annotated mass lesion regions in the training set. It can be seen from the figure that although the size of the occupied areas varies, the distribution is concentrated in the [64, 128] interval, and the distribution conforms to the normal distribution. Therefore, it is suitable to set the scale and aspect ratio of the anchor to 64(1) and 128(1). The scale of 256 and 512 is too large to draw the lesion area, which also verifies the choice of the anchor in the above table.
Obviously, the IoU is a key parameter, and its size will directly affect the fault tolerance performance of the detection network model in actual samples. When the threshold of IoU is set higher, it indicates that the network has stricter requirements for regional overlap. The ROC curve, also known as the susceptibility curve, is a curve with the FPR value as the abscissa and the TPR as the ordinate. It is often used to describe the pros and cons of the classification model. The size of the area under the curve (AUC) directly reflects the performance of the model, and the size of the AUC is positively related to the quality of the model. Fig\ref{fig:wce_detect_roc} plots the ROC curves under different IoU thresholds. It can be seen from the chart that the threshold value range is [0,1], and the AUC value is 97.2$\%$. After comprehensive analysis, it is decided to use the IoU threshold of 0.6, because the detection performance of the network is the best at this time.
Tab\ref{tab:detection_networks_anchors} shows the combination when the anchor's scale parameter is set to 64, the aspect ratio is (1, 0.5, 2) and when the anchor's scale parameter is set to 128 and the aspect ratio parameter is set to (1) The performance in the patient test, at this time, the anchor parameter combination is abbreviated as 64³+128². Tab\ref{tab:detection_networks_anchors} shows the statistical results of the Scale parameter of 64, the aspect ratio parameter of (1) and the Scale of 128 and the aspect ratio of (1). The parameter combination at this time can be abbreviated as 64²+128². The statistical data in these two tables are derived from the results of the 7 patients with the corresponding anchor parameter combination. Based on the performance and index comparisons in tab\ref{tab:detection_networks_anchors}, fig\ref{fig:wce_length_width_select} and tab\ref{tab:detection_networks_result_diffent_anchors}, it is determined that the scale and aspect ratio parameters of the Anchor in the improved Faster R-CNN are set to 64(1) and 128(1).\\
\begin{table}[!hbp]
    \footnotesize
    \centering
    \caption{Test results of space-occupying patients under different anchors (part) and doffernt patient, and A to G are the 7 patients used in this phase of the experiment.}
    \label[]{tab:detection_networks_result_diffent_anchors}
    \resizebox{\linewidth}{!}{
    \begin{tabular}{ccccccccccccc} \toprule
        & \multicolumn{6}{c}{Space-occupying patient test results under 64³+128² anchor}      & \multicolumn{6}{c}{Space-occupying patient test results under 64²+128² anchor}      \\ \midrule
        & \textbf{R}\_sen(\%) & \textbf{R}\_spe(\%) & R\_acc(\%) & \textbf{I}\_sen(\%) & \textbf{I}\_spe(\%) & \textbf{I}\_acc(\%) & \textbf{R}\_sen(\%) & \textbf{R}\_spe(\%) & \textbf{R}\_acc(\%) & \textbf{I}\_sen(\%) & \textbf{I}\_spe(\%) & \textbf{I}\_acc(\%) \\
        A & 94.62      & 8.43       & 91.6       & 95.8       & 7.6        & 92.42      & 98.7       & 8.3        & 91.78      & 98.6       & 7.2        & 92.86      \\
        B & 95.4       & 11.58      & 88.43      & 90.86      & 11.02      & 88.99      & 95.4       & 11         & 89.03      & 95.4       & 10.4       & 89.64      \\
        C & 93.38      & 5.21       & 94.79      & 93.38      & 4.5        & 95.45      & 100        & 5.1        & 95.01      & 100        & 4.8        & 95.21      \\
        D & 98.15      & 9.11       & 83.95      & 97.81      & 11.11      & 89.08      & 98.6       & 8.96       & 91.2       & 98.48      & 8.07       & 92.07      \\
        E & 94.44      & 8.87       & 91.12      & 94.44      & 8.85       & 91.15      & 94.44      & 6.1        & 94         & 94.44      & 4.2        & 95.78      \\
        F & 97.17      & 6.55       & 95.82      & 97.46      & 6.72       & 93.59      & 98.38      & 4.2        & 95.85      & 98.37      & 3.72       & 96.31      \\
        G & 78.33      & 7.5        & 92.48      & 81.67      & 6.49       & 93.51      & 98.33      & 7.51       & 92.51      & 98.33      & 7.41       & 92.61         \\ \bottomrule
    \end{tabular}
}
\end{table}
\noindent\textbf{Comparative experiment result in this section.}
The improved Faster R-CNN\textcolor{red}* in tab\ref{tab:detection_networks_result_A_O} shows the results when redundant data such as residuse and bubbles are not added. From the results shown in these figures, it can be found that the data containing residues and bubbles will greatly interfere with the final experiment. The performance of the lesion detection scheme. Since the number of redundant images in the WCE image sequence of patient A (determined by the clinician as the type of occupation lesion) is less than 15$\%$ of the total images. However, the number of redundant images in the image sequence of patient O (the type of non-occupying lesions determined by clinicians) accounts for 15$\%$-30$\%$ of the entire sequence. When redundant data is added to the training sample, the error of Faster R-CNN\textcolor{red}* is improved. There is a significant drop in the detection rate indicator.
\begin{table}[!hbp]
    \footnotesize
    \centering
    \caption{The detection results of cases A and O at the RPN level and the images level in each contrastive lesion detection network.}
    \label[]{tab:detection_networks_result_A_O}
    \resizebox{\linewidth}{!}{
    \begin{tabular}{ccccccccccc} \toprule
        & \multicolumn{6}{c}{patient A}                                               & \multicolumn{4}{c}{patient O}                     \\ \midrule
        \textbf{}                     & \textbf{R}\_acc(\%) & \textbf{R}\_sen(\%) & \textbf{R}\_spe(\%) & \textbf{I}\_acc(\%) & \textbf{I}\_sen(\%) & \textbf{I}\_spe(\%) & \textbf{R}\_acc(\%) & \textbf{I}\_acc(\%) & \textbf{R}\_spe(\%) & \textbf{I}\_spe(\%) \\
        Original version Faster R-CNN & 80.59      & 88.7       & 19         & 80.69      & 94.4       & 19.5       & 69.17      & 69.25      & 30.83      & 30.75      \\
        Faster R-CNN-Res2net          & 75.07      & 91.88      & 75.07      & 75.15      & 91.36      & 25         & 68.63      & 68.7       & 31.37      & 31.3       \\
        Improved Faster R-CNN         & \textbf{91.78}      & \textbf{98.7}       & \textbf{8.3}        & \textbf{92.86}      & \textbf{98.6}       & \textbf{7.2}        & \textbf{80.01}      & \textbf{80.52}      & \textbf{19.99}      & \textbf{19.48}      \\
        Improved Faster R-CNN\textcolor{red}*        & 48.11      & 98.7       & 51.99      & 74.33      & 98.7       & 25.92      & 54.12      & 54.12      & 45.88      & 45.88     \\ \bottomrule
    \end{tabular}
}
\end{table}
Analysis from the overall patient case level means that the detection model predicts whether the case label of the overall patient is the same as the patient label labeled by the doctor from the RPN and picture levels. If the overlapping area with the manually annotated area exceeds 60$\%$, the area is judged as space-occupying; if the patient's image predicts the existence of space-occupying lesions, the patient's disease type is judged as space-occupying. The statistical content includes the difference between the categories framed by the RPN module and the actual patient lesions, and the detection effect at the overall case level at this stage. Tab\ref{tab:detection_networks_result_3} plots the detection results of each lesion detection network on the current test data at the RPN level and the picture level. The test data consists of the WCE image sequences of 21 test patients. It can be found from the figure that the indicators of the comparison network Faster R-CNN-FPN are the lowest in the detection network, and the three indicators cannot meet the practical application in RPN and Image level; while the remaining three networks can obviously achieve a high level of overall data level. Sensitivity, but the false detection rate of the improved network Faster R-CNN is better than other networks, and the data of indicators such as accuracy rate is the best.
\begin{table}[!hbp]
    \footnotesize
    \centering
    \caption{Comparison of the results of each occupancy detection network on the current test data at the RPN level(\textbf{R}) and the Image level(\textbf{I}).acc, sen and spe refer to accuracy, sensitivity and specificity}
    \label[]{tab:detection_networks_result_3}
    \resizebox{\linewidth}{!}{
    \begin{tabular}{ccccccc} \toprule
        & \textbf{R}\_acc(\%) & \textbf{R}\_sen(\%) & \textbf{R}\_spe(\%) & \textbf{I}\_acc(\%) & \textbf{I}\_sen(\%) & \textbf{I}\_spe(\%) \\ \midrule
        Original Faster R-CNN & 55         & 100        & 69.23      & 60         & 71.42      & 28.57      \\
        Imitation Faster R-CNN  & 50         & 100        & 76.92      & 45         & 85.71      & 76.92      \\
        Improved Faster R-CNN & \textbf{80.95}      & \textbf{100}        & \textbf{28.55}      & \textbf{85.71}      & \textbf{100}        & \textbf{21.42}      \\
        Faster R-CNN-FPN     & 55         & 85.71      & 61.53      & 50         & 71         & 61.54     \\ \bottomrule
    \end{tabular}
}
\end{table}
So far, the experimental results shown in the paper are all data-level results, and now the results of image visualization are shown. The first row shows the labeling and classification effect of the currently used occupancy detection network model on the same test image, and the next row shows the results of different improved detection networks tested on different images. First of all, comparing and analyzing fig\ref{fig:wce_detect_result_image}, it can be seen that the fine-tuned network annotation occupies the most accurate place, while both the traditional Faster R-CNN and the imitation Faster R-CNN network have false detections, and the Faster R-CNN-FPN network is not detected. To the lesion, that is, missed detection; finally, from the second row, it can be found that the detection effect of the traditional Faster R-CNN's false detection and fine-tuning network is improved.

Through the experimental verification, the improved Faster R-CNN can screen out thespace-occupying lesion image in the WCE sequence, which can achieve the expected effect of the experiment and prove that the detection scheme is feasible. Secondly, the improved Faster R-CNN has obvious improvement in both data display level and image visualization level. The improved network can improve the accuracy index compared with the traditional detection network, and the sensitivity index is unchanged, but the false detection rate index decreases significantly. Next, compared with other contrast algorithms imitation Faster R-CNN and Faster R-CNN-FPN, the accuracy, sensitivity and false detection rate are more prominent. However, there are shortcomings after the end of this stage: the false positive rate is very high. In practical application, there are many non-space-occupying images in the space-occupying data considered by the network. The process of re-screening is still a re-reading and inefficient work for doctors. Therefore, there is no real sense to detect the space-occupying lesion image from the WCE image sequence.
\subsection{lesion recognition network}
The test set currently used in the testing phase uses 21 cases, including 7 space-occupying, 7 bleeding and ulcer, and 7 normal. The lesion recognition phase of the paper draws summary results at the overall patient level for 21 cases in tab\ref{tab:recognition_networks_result_1}. The table reports the experimental results at this stage from three perspectives: the overall patient case level, the specific patient level, and the overall patient image level. Each level also involves the detection and screening effects at the RPN level and the picture level. Among them, WCE-RFNet is a mass lesion recognition network based on multi-scale feature fusion proposed in this paper, and VGG16, DPN\cite{chen2017dual,zhu2018deeplung}, and ResNet50 are common classification networks. It can be seen from the figure that the Faster R-CNN network modified on the data has a high false detection rate in the three indicators, and after the classification network, there is a certain degree of decline, and the accuracy index remains basically unchanged. However, the sensitivity and false detection rate indicators are different: although the DPN and VGG16 calibration models can reduce the false detection rate, the data statistics at the image level also reduce the sensitivity index, resulting in missed detection of actual data; The robustness of the calibration network with the residual module is great, which ensures high sensitivity and reduces the false detection rate.
\begin{table}[H]
    \footnotesize
    \centering
    \caption{Summary results (image level) of lesion recognition network in 21 patients.}
    \label[]{tab:recognition_networks_result_1}
    \begin{tabular}{cccc} \toprule
        & \textbf{I}\_acc(\%) & \textbf{I}\_sen(\%) & \textbf{I}\_spe(\%) \\ \midrule
        Improved Faster R-CNN & 85.71      & 85.71      & 21.43      \\
        DPN                   & 71.42      & 71.43      & 28.57      \\
        VGG16                 & 76.19      & 85.71      & 28.57      \\
        ResNet50              & 85.71      & 85.71      & 14.29      \\
        WCE-RFNet(ours)             & \textbf{90.48}      & \textbf{100}        & \textbf{7.14}      \\ \bottomrule
    \end{tabular}
\end{table}
Currently, the test sets before and after are added in the testing stage, and experiments are performed in the stage of detection of occupied lesions and the stage of identification of occupied lesions. In this paper, a result table such as tab\ref{tab:recognition_networks_result_2} is produced from the two parts before and after the addition at the overall patient level. It can be seen from the table: First, the effects of the lesion detection stage and the lesion identification stage performed well before and after the increase of the test set samples. The final detection index sensitivity, false positive rate and accuracy were 100$\%$, 6.67$\%$ and 94.56$\%$, which was significantly reduced from the data of false positive indicators; secondly, after the increase of test set samples, the false positive rate of the model in the lesion detection stage decreased by 8.52, and the final accuracy rate of the model in the lesion identification stage increased by 4.08. It can be seen from the overall patient level that the scheme proposed in this paper is effective.
\begin{table}[]
    \footnotesize
    \centering
    \caption{Comparison of index data in each stage of the whole patient before and after adding test data set(patient leverl).}
    \label[]{tab:recognition_networks_result_2}
    \begin{tabular}{ccccc} \toprule
        \multirow{2}{*}{} & \multicolumn{2}{c}{before increase}                                                                                                           & \multicolumn{2}{c}{after increase}                                                                                                            \\ \midrule
        & \begin{tabular}[c]{@{}c@{}}lesion\\ detection\\ network\end{tabular} & \begin{tabular}[c]{@{}c@{}}lesion\\ recognition\\ network\end{tabular} & \begin{tabular}[c]{@{}c@{}}lesion\\ detection\\ network\end{tabular} & \begin{tabular}[c]{@{}c@{}}lesion\\ recognition\\ network\end{tabular} \\
        sen(\%)           & 100                                                                  & 100                                                                    & 100                                                                  & \textbf{100}                                                           \\
        spe(\%)           & 21.42                                                                & 7.14                                                                   & \textbf{12.90}                                                                & \textbf{6.67}                                                          \\
        acc(\%)           & 33.33                                                                & 90.48                                                                  & \textbf{91.1}                                                                 & \textbf{94.56}                                                          \\ \bottomrule
    \end{tabular}
\end{table}
\subsection{overall detection framwork}
\noindent\textbf{The experimental results of this paper.}
After cascading the lesion detection network and identification, the final lesion detection framework model of this paper is composed. The model summarizes the experimental results in tab\ref{tab:overall_detect_result_21} at the overall picture level for the test results of this batch of 21 patients. It can be seen from the table that in this batch of test data consisting of 21 patients, the values of sensitivity, false positive rate and accuracy rate can be obtained after one-stage processing are 98.28$\%$, 6.3$\%$ and 93.71$\%$; The values that can be obtained after one-stage processing are 97.95$\%$, 5.45$\%$ and 94.57$\%$. Numerically, it was found that the sensitivity decreased by 0.33, the false positive rate decreased by 0.85, and the accuracy increased by 0.85; the decrease in sensitivity at this time was caused by too few test samples.
\begin{table}[]
    \footnotesize
    \centering
    \caption{Comparison of indicators at the overall image level under the test set of 21 patients at each stage.}
    \label[]{tab:overall_detect_result_21}
    \begin{tabular}{cccc} \toprule
        & sen(\%) & spe(\%) & acc(\%) \\ \midrule
        lesion detection network   & 98.28   & 6.3     & 93.71   \\
        lesion recognition network & 97.95   & 5.45    & 94.56  \\ \bottomrule
   \end{tabular}
\end{table}
In the case of the test data set composed of 45 patients, the overall detection framework of this paper summarizes tab\ref{tab:overall_detect_result_45} at the overall picture level, and tab\ref{tab:overall_detect_result_all} is a summary of the confusion matrix of the test set before and after the addition. It can be seen from the two tables: in this batch of test data composed of 45 patients, the values of sensitivity, false positive rate and accuracy rate indicators can be obtained at the picture level after processing in the lesion detection stage are 98.75$\%$, 7.41$\%$ and 92.60$\%$; and after the treatment of the lesion identification stage, the index values became 98.75$\%$, 5.62$\%$ and 94.39$\%$. From the numerical value, it can be found that after increasing the test sample, the sensitivity can be guaranteed to remain unchanged, the false positive rate is decreased by 1.79, and the accuracy rate is increased by 1.79. It shows that the detection accuracy of the model can be improved while the false positive rate is reduced in the stage of lesion identification.
\begin{table}[]
    \footnotesize
    \centering
    \caption{The added test data set compares the indicator data at all levels of the overall picture.}
    \label[]{tab:overall_detect_result_45}
    \resizebox{\linewidth}{!}{
    \begin{tabular}{ccccc} \toprule
        \multirow{2}{*}{} & \multicolumn{2}{c}{\begin{tabular}[c]{@{}c@{}}lesion \\ detection\\ network\end{tabular}}                   & \multicolumn{2}{c}{\begin{tabular}[c]{@{}c@{}}lesion\\ recognition\\ network\end{tabular}}                  \\ \midrule
        & \begin{tabular}[c]{@{}c@{}}RPN\\ level\end{tabular} & \begin{tabular}[c]{@{}c@{}}Image\\ level\end{tabular} & \begin{tabular}[c]{@{}c@{}}RPN\\ level\end{tabular} & \begin{tabular}[c]{@{}c@{}}Image\\ level\end{tabular} \\
        sen(\%)           & 98.81                                               & 98.75                                                 & /                                                   & \textbf{98.75}                                                 \\
        spe(\%)           & 7.43                                                & 7.41                                                  & /                                                   & \textbf{5.62}                                                  \\
        acc(\%)           & 92.57                                               & 92.60                                                 & /                                                   & \textbf{94.39}                                                \\ \bottomrule
    \end{tabular}
}
\end{table}
\begin{table}[!hbp]
    \footnotesize
    \centering
    \caption{Confusion matrices at each stage at the overall patient picture level before and after test dataset augmentation.}
    \label[]{tab:overall_detect_result_all}
    \resizebox{\linewidth}{!}{
    \begin{tabular}{ccccc} \toprule
        \multirow{4}{*}{\begin{tabular}[c]{@{}c@{}}\\ \\Lesion detection network\\ (not increased)\end{tabular}}   &                                                                & \begin{tabular}[c]{@{}c@{}}Positive\\ (Actual)\end{tabular} & \begin{tabular}[c]{@{}c@{}}Negative\\ (Actual)\end{tabular} & Total     \\ \midrule
        & \begin{tabular}[c]{@{}c@{}}Positive\\ (Predicted)\end{tabular} & 2,873                                                       & 53,822                                                      & 56,695    \\
        & \begin{tabular}[c]{@{}c@{}}Negative\\ (Predicted)\end{tabular} & 50                                                          & 799,617                                                     & 799,667   \\
        & Total                                                          & 2,923                                                       & 853,499                                                     & 856,362   \\
        \multirow{4}{*}{\begin{tabular}[c]{@{}c@{}}\\ \\Lesion recognition network\\ (not increased)\end{tabular}} &                                                                & \begin{tabular}[c]{@{}c@{}}Positive\\ (Actual)\end{tabular} & \begin{tabular}[c]{@{}c@{}}Negative\\ (Actual)\end{tabular} & Total     \\  \midrule
        & \begin{tabular}[c]{@{}c@{}}Positive\\ (Predicted)\end{tabular} & 2,863                                                       & 46,516                                                      & 49,379    \\
        & \begin{tabular}[c]{@{}c@{}}Negative\\ (Predicted)\end{tabular} & 60                                                          & 806,923                                                     & 806,983   \\
        & Total                                                          & 2,923                                                       & 853,439                                                     & 856,362   \\
        \multirow{4}{*}{\begin{tabular}[c]{@{}c@{}}\\ \\lesion recognition network\\ (increased)\end{tabular}}     &                                                                & \begin{tabular}[c]{@{}c@{}}Positive\\ (Actual)\end{tabular} & \begin{tabular}[c]{@{}c@{}}Negative\\ (Actual)\end{tabular} & Total     \\ \midrule
        & \begin{tabular}[c]{@{}c@{}}Positive\\ (Predicted)\end{tabular} & 2,221                                                       & 61,138                                                      & 63,359    \\
        & \begin{tabular}[c]{@{}c@{}}Negative\\ (Predicted)\end{tabular} & 28                                                          & 1,026,521                                                   & 1,026,549 \\
        & Total                                                          & 2,249                                                       & 1,087,659                                                   & 1,089,908 \\
        \multirow{4}{*}{\begin{tabular}[c]{@{}c@{}}\\ \\ Overall detection network\\ \\ (increased)\end{tabular}}   &                                                                & \begin{tabular}[c]{@{}c@{}}Positive\\ (Actual)\end{tabular} & \begin{tabular}[c]{@{}c@{}}Negative\\ (Actual)\end{tabular} & Total     \\ \midrule
        & \begin{tabular}[c]{@{}c@{}}Positive\\ (Predicted)\end{tabular} & 5,084                                                       & 107,654                                                     & 112,738   \\
        & \begin{tabular}[c]{@{}c@{}}Negative\\ (Predicted)\end{tabular} & 88                                                          & 1,833,444                                                   & 1,833,532 \\
        & Total                                                          & 5,172                                                       & 1,941,098                                                   & 1,946,270 \\ \bottomrule
    \end{tabular}
}
\end{table}

\noindent\textbf{Compared with other methods.}
In this paper, the detection structure as shown in Tab\ref{tab:paper_detection_stage_training_all} is finally adopted as the detection framework of the gastrointestinal lesions in the WCE sequence based on feature fusion. A test dataset consisting of 45 patients was tested. The final statistics are shown in Tab\ref{tab:paper_detection_stage_training_all} below, which is the data comparison of the overall test data set on the indicators at all levels on the overall picture. It can be seen from the table that the final result of the detection method in this paper is to obtain the sensitivity, The values of the false positive rate and accuracy rate indicators were 98.75$\%$, 5.62$\%$ and 94.39$\%$, and the values of the sensitivity, false positive rate and accuracy rate indicators at the overall patient level were 100$\%$, 6.67$\%$ and 94.56$\%$.
\begin{table}[]
    \footnotesize
    \centering
    \caption{The overall test data set comparison of each indicator data at all image levels with this paper.}
    \label[]{tab:paper_detection_stage_training_all}
    \begin{tabular}{ccc} \toprule
        & \begin{tabular}[c]{@{}c@{}}lesion detection\\ network\end{tabular} & \begin{tabular}[c]{@{}c@{}}lesion recognition\\ network\end{tabular} \\ \midrule
        sen(\%) & 98.28                                                              & 98.30                                                                \\
        spe(\%) & 7.41                                                               & 5.55                                                                 \\
        acc(\%) & 93.71                                                              & 94.46                                                               \\ \bottomrule
    \end{tabular}
\end{table}
The paper also introduces common detection networks for image detection of gastrointestinal mass lesions based on WCE images, such as YOLOv4, SSD, M2Det, EfficientDet and CenterNet. Compared with the network statistics used in this paper, the detection comparison table at the overall picture level is obtained in Tab\ref{tab:detection_stage_testing_compare}. From the table, we can clearly and directly see the advantages of the model. The scheme proposed in this paper is efficient and excellent. Among them, M2Det, CenterNet, EfficientDet and YOLOv4 are the first applications to fully use the lesion detection network proposed on the natural dataset to train and test on the WCE dataset. The network model proposed in this paper can meet the actual needs and scenarios, and shows high sensitivity and low false detection rate in many algorithms, which is a must have in the field of lesion detection in medical image processing.
\begin{table}[]
    \footnotesize
    \centering
    \caption{The overall test data set is the indicator data on the overall picture in different detection networks.}
    \label[]{tab:detection_stage_testing_compare}
    \begin{tabular}{cccc} \toprule
        & sen(\%) & spe(\%) & acc(\%) \\ \midrule
        M2det     & 96.04   & 7.62    & 92.38   \\
        CenterNet   & 83.64   & 29.2    & 70.85   \\
        EfficientDet & 85.25   & 11.58   & 85.25   \\
        YOLOv4       & 92.36   & 10.61   & 89.4    \\
        SSD\cite{aoki2019automatic}          & 97.25   & 8.9     & 91.12   \\
        Lan\cite{lan2019deep}          & 78.16   & 0.08    & /       \\
        Ours         & \textbf{98.75}   & \textbf{5.62}    & \textbf{94.39}   \\ \bottomrule
    \end{tabular}
\end{table}

\begin{figure}[!hbtp]
    \centering
    \includegraphics[width=\linewidth]{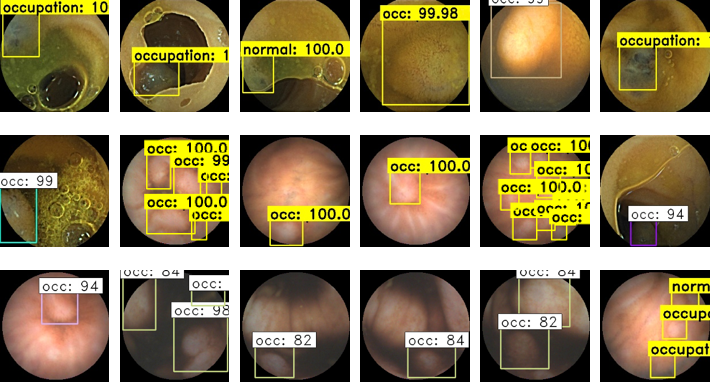}
    \caption{The detection results of the detection framework in this paper in tumors and polyps.}
    \label{fig:omom_wce_paper_result}
\end{figure}

Fig\ref{fig:omom_wce_paper_result} shows the test results of the proposed mass lesion detection network TMFnet on different images containing polyps and tumors of different sizes. From the model annotation results in the figure, it can be shown that the detection scheme proposed in this paper can detect mass-occupying lesions of different scales, and the effect is considerable, and it can adapt to the clinical digestive tract WCE sequence of the real scene.

The core algorithm of a successful medical detection system often determines the response and user experience of the system. The algorithm model proposed in this paper can detect 4 or 5 patients during the normal working time of the day, which is better than the work efficiency of gastroenterologists. The detection efficiency of the algorithm in a WCE image sequence is 8FPS (Frames Per Second). , including 3FPS of the lesion identification network. This shows that the algorithm has high detection efficiency.

\section{Discussion and Conclusion}
In the stage of occupancy detection, the experimental scheme and network structure proposed in the experiment can be found in the comparison with other networks: the network structure can more comprehensively learn the feature set of the WCE training set image, and obtain the appropriate weight of the network, so it can In the network testing stage, it shows high accuracy, high sensitivity, and high false detection rate; it also shows on another level: the deeper the network structure, the stronger the automatic learning ability of features, and the recognition rate of the model also increases. The comparison network Faster R-CNN-FPN and Imitation Faster R-CNN did not learn enough times in the same batch of training data sets, and both have great room for improvement in the experiment. Especially from fig\ref{fig:wce_detect_result_image}, it can be seen that the localization and classification of the test phase in the comparison network are not as good as the improved Faster R-CNN. 

In the lesion recognition stage, it can be seen that the four recognition networks used in the paper can complete image recognition to a certain extent, but the Faster R-CNN network modified on the data has a high false detection rate in the three indicators. After the classification network, there is a certain degree of decline, and the accuracy index can generally be maintained. However, the sensitivity and false detection rate indicators are different: although DPN and VGG16 reduce the false detection rate at the image level, they also reduce the sensitivity indicator. When the excellent model weights of the two are finally cascaded in the paper, it is found after the increased test samples that the scheme exhibits high sensitivity and accuracy, and more importantly, low false detection rate. The final experimental results of the scheme are: the values of the sensitivity, false positive rate and accuracy rate at the picture level are 98.75$\%$, 5.62$\%$ and 94.39$\%$, and the sensitivity, false positive rate and accuracy rate are obtained at the overall patient level The values of the indicators are 100$\%$, 6.67$\%$ and 94.56$\%$. 

It can be seen from the final experimental results that this scheme is expected to be able to effectively screen out pictures with space-occupying lesion information from the digestive tract sequence, which can greatly reduce the workload of doctors and improve work efficiency.It can be seen from the fig\ref{fig:omom_wce_paper_result} that the lesion detection method of the two-stage multi-scale feature-fusion network TMFNet proposed in this paper performs well in the screening of gastrointestinal tumors and polyps, especially for small polyps. At the same time, the scheme proposed in this paper is compared with other detection methods and found that it has the advantages of irreplaceable (State-Of-The-Art).
{\small
\bibliographystyle{ieee_fullname}
\bibliography{wcebib}
}

\end{document}